\documentclass{article}

\usepackage{arxiv}

\usepackage[utf8]{inputenc} % allow utf-8 input
\usepackage[T1]{fontenc}    % use 8-bit T1 fonts
\usepackage{hyperref}       % hyperlinks
\usepackage{url}            % simple URL typesetting
\usepackage{booktabs}       % professional-quality tables
\usepackage{amsfonts}       % blackboard math symbols
\usepackage{nicefrac}       % compact symbols for 1/2, etc.
\usepackage{microtype}      % microtypography
\usepackage{lipsum}		% Can be removed after putting your text content
\usepackage{graphicx}
\usepackage{doi}
\usepackage{cite}

\usepackage{lineno}
\usepackage{algorithm}
\usepackage{algpseudocode}
\usepackage[title]{appendix}
\usepackage{amsmath}

\title{Efficient reference-less transmission matrix retrieval for a multimode fiber using fast Fourier transform}

\author{ {Jingshan Zhong}{\dag} \\
	Research Center for Humanoid Sensing\\
    Zhejiang Lab\\
    Hangzhou 311100, China\\
	\texttt{zhongjingshan@hotmail.com} \\
	\And
	{Zhong Wen{\dag}, Quanzhi Li, Qilin Deng,  and Qing Yang}\thanks{Please contact Qing Yang: \texttt{qingyang@zju.edu.cn}.} \\
	State Key Laboratory of Extreme Photonics and Instrumentation\\
    College of Optical Science and Engineering\\
    International Research Center for Advanced Photonics\\
    Zhejiang University, Hangzhou 310027, China\\
	\texttt{\{21730014,12130083,22030075,qingyang\}@zju.edu.cn} \\
}

\renewcommand{\shorttitle}{ reference-less TM retrieval using FFT}

\begin{document}
\maketitle

\begin{abstract}
	Transmission matrix (TM) linearly maps the incident and transmitted complex fields, and has been used widely due to its ability to characterize scattering media. It is computationally demanding to reconstruct the TM from intensity images measured by a reference-less experimental setup. Removing reference beam for interference gains the advantage of simple experimental setup. However, the long computational time still limits its practical application. We propose an efficient reference-less TM retrieval method for multimode fiber (MMF). Our method adopts a data acquisition scheme which employs Fourier transform matrix in the design of the incident fields. We develop a nonlinear optimization algorithm to solve the TM retrieval problem in a parallel manner. The data acquisition scheme allows the algorithm to be implemented with fast Fourier transform (FFT), and hence achieves great efficiency improvement. Further, our method acquires intensity images at a defocus plane and correct the error of relative phase offset of TM recovered from the intensity images measured at one fixed plane. We validate the proposed TM retrieval method with both simulations and experiments. By using FFT, our TM retrieval algorithm achieves 1200x speed-up in computational time, and recovers $2286 \times 8192$ TM of a 0.22 NA and $50 \ \mu m$ diameter MMF with 124.9 seconds by a computer of 32 CPU cores. With the advantages of efficiency and the correction of phase offset, our method paves the way for the application of reference-less TM retrieval in real practice.
\end{abstract}

% keywords can be removed

%%%%%%%%%%%%%%%%%%%%%%%%%%  body  %%%%%%%%%%%%%%%%%%%%%%%%%%
\section{Introduction}

When light transports through complex media, such as the fog, biological tissue, or MMF, scattering usually scrambles the incident field into random patterns. Fortunately, a TM fully characterizes the light transmission property of static scattering medium. The TM describes the linear relationship of the incident and transmitted complex fields. By using the TM, one can manipulate the incident complex field to generate desired output patterns~\cite{popoff2010measuring} or inverse the scattering process to retrieve the information of the incident complex field from measured speckle patterns~\cite{popoff2010image}.

Techniques based on TM have shown potential in a broad range of applications, including focusing~\cite{vellekoop2007focusing, popoff2010measuring, papadopoulos2012focusing, cui2011parallel}, imaging~\cite{popoff2010image,vcivzmar2012exploiting,choi2012scanner}, optical communications~\cite{carpenter2015observation}, optical computing~\cite{matthes2019optical,gong2019optical}, and quantum networks~\cite{leedumrongwatthanakun2020programmable}. Imaging through a MMF of hundreds micron diameter is possible by exploiting the property of TM. By measuring the TM of the MMF beforehand, one can control the incident field by digital micromirror devices (DMD) or spatial light modulators (SLM) and generate fast scanning light patterns on the imaging sample. The light could be manipulated into diverse forms, such as 2D foci~\cite{popoff2010measuring}, 3D foci~\cite{wen2020fast}, light sheet~\cite{ploschner2015multimode}, and even user-specified patterns~\cite{vcivzmar2011shaping}. The MMF imaging techniques achieve capabilities of fluorescence imaging~\cite{ploschner2015multimode,turtaev2018high}, optical tweezers~\cite{bianchi2012multi}, and remote depth sensing~\cite{stellinga2021time}. Another example is the application of TM in high capacity optical communications. It allows investigation of the principle modes of a MMF~\cite{carpenter2015observation}, and study of efficient data transmission~\cite{gong2019optical}. 

However, the acquisition of a TM could pose technical limitations which prevents wide application of the TM techniques. The TM contains complex-value entries, whose phase cannot be directly measured by a camera. It requires interferometric setup to measure the lost phase information with an external reference beam~\cite{choi2011overcoming,vcivzmar2011shaping,li2021compressively}. It needs complicated experimental setup and suffers from instability. Besides, it is difficult to obtain an external reference to measure TM of kilometers long fiber in the application of optical communication. By removing the external reference beam, a portion of the modulation device can be set as the internal reference beam which co-propagates through the medium~\cite{popoff2010measuring,conkey2012high,yoon2015measuring}. However, it reduces the number of effective modulation modes, and results in speckle reference beam which causes missing points on TM measurements~\cite{stellinga2021time}. Methods based on Bayesian approach~\cite{dremeau2015reference,deng2018characterization}, Gerchberg-Saxton~\cite{huang2021generalizing}, semi-definite programming~\cite{n2017controlling}, Kalman filter~\cite{huang2020retrieving}, gradient descent based method~\cite{cheng2022non} have been proposed to computationally recover the TM from transmitted speckle intensity images. These methods have advantage of removing the requirement of either external or internal reference. Since the TM typically has very large size, the computational time of these methods could be hours~\cite{huang2021generalizing}, limiting its practical use. 

Another common issue is that the TM acquired by the methods of internal reference ~\cite{popoff2010measuring,conkey2012high,yoon2015measuring} or computational recovery~\cite{dremeau2015reference,deng2018characterization,huang2021generalizing,n2017controlling,huang2020retrieving} have error of phase offset compared to the true TM~\cite{bianchi2012multi}. The former generate a speckle reference field of unknown phase while the latter use intensity images measured at one fixed camera plane, which misses the information of relative phases between different pixels. The transmitted complex field predicted by multiplication of the acquired TM and incident complex field has correct amplitudes but wrong phases. It causes failure of generating 3D light patterns, such as 3D foci or light sheet, which is essential in volumetric imaging~\cite{wen2020fast} and light sheet imaging by MMF~\cite{ploschner2015multimode}.

\begin{figure}[htpb]
	\centering
	\includegraphics[width=0.95\linewidth]{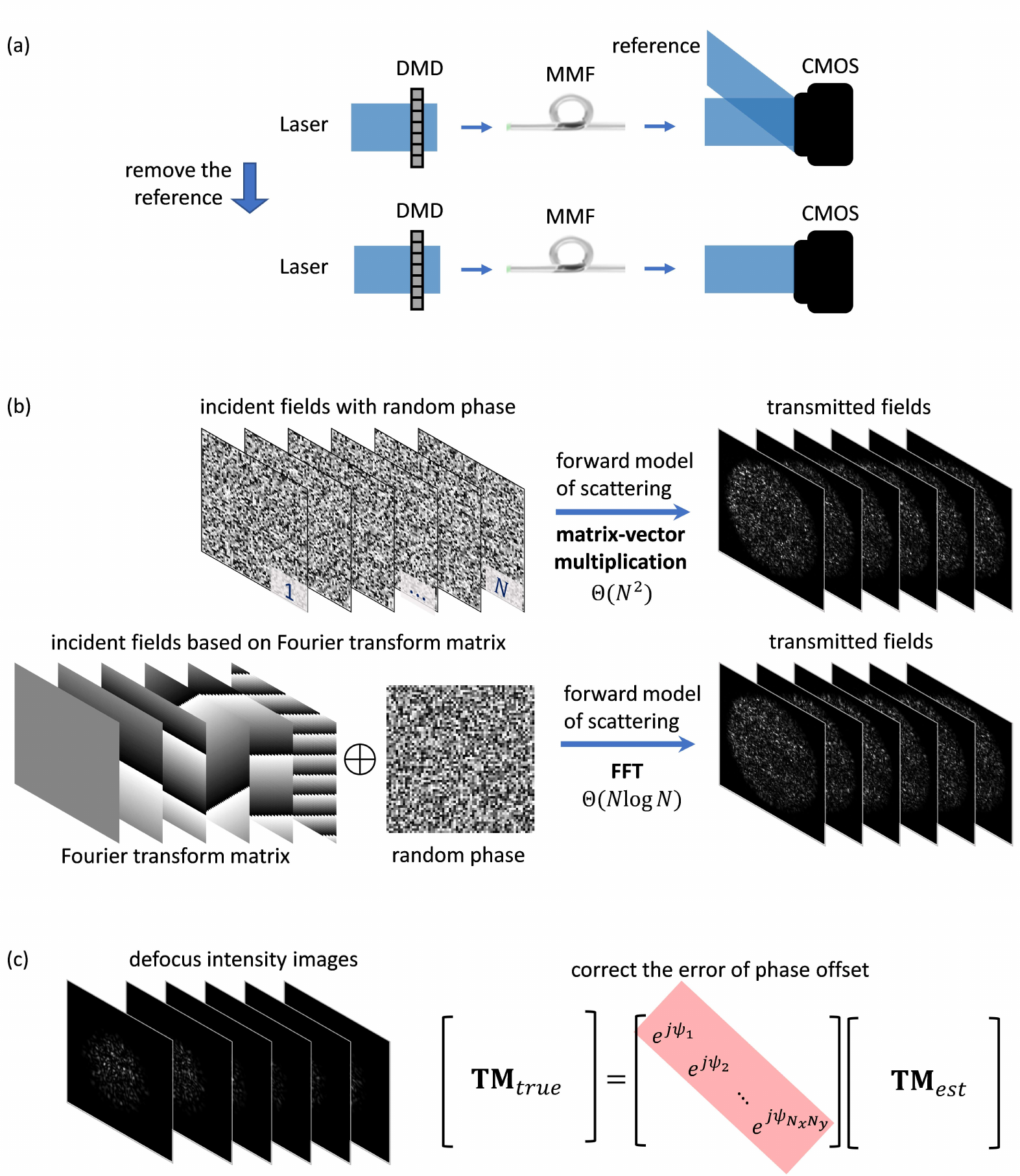} 
	\caption{ TM retrieval with FFT and phase correction from intensity measurements without reference. (a) Comparison of data acquisition between our method and the reference-based methods. The reference-based methods interferometrically measure the complex fields with a reference beam while our method takes only intensity without any reference, leading to a simpler experiment. (b) Computational efficiency improvement by using FFT. In the case that the incident fields are directly generated with random phases, the forward model of scattering has to be computed by matrix-vector multiplication. Our method designs the incident fields based on Fourier transform matrix. Thus, the forward model of scattering can be computed by FFT, which significantly improves the computaional efficiency. It also allows the inverse algorithm of TM retrieval to be implemented with FFT. (c) Correction of the error of phase offset by using defocus intensity images. The estimated TM from the intensity images measured at one defocus plane has the error of phase offset. Our algorithm corrects the phase errors by using the defocus intensity images.}
	\label{fig:scatteringbyFFT}
\end{figure}

In this work, we propose an efficient method to recover the TM from a reference-less experimental setup. It removes the reference beam which is used to measure amplitude and phase of the transmitted complex fields (Fig.~\ref{fig:scatteringbyFFT}(a)). Compared with the reference-based techniques, the reference-less experimental setup is simpler and considerably more stable. First,  we design a probing matrix based on Fourier transform matrix. It is applied on a phase modulator to generate incident complex fields. The intensity images of the corresponding transmitted complex fields are measured by the reference-less experimental setup for TM retrieval. We develop a nonlinear optimization algorithm to solve the TM from the intensity images. Compared to the random probing matrix, our probing matrix allows the inverse algorithm to be implemented with FFT (Fig.~\ref{fig:scatteringbyFFT}(b)), which greatly reduces the computational complexity. Second, we correct the error of phase offset in recovered TM due to the intensity images measured at a fixed measurement plane (Fig.~\ref{fig:scatteringbyFFT}(c)). Our method measures a set of intensity images at a defocus plane and develops an algorithm to recover the phase offset from the defocus intensity images. We verify the proposed method by simulations and experiments. The simulation shows the TM retrieval algorithm with FFT has 1200x speed-up in computational time compared to that of the TM retrieval algorithm without FFT. The proposed TM retrieval method recovers TM for a MMF of 0.22 NA and $50 \ \mu m$ diameter with 124.9 seconds. We build the experimental setup by using MMF, recover the TM with experimental data sets, and verify the proposed methods by evaluating the foci in both 2D and 3D.

 %Computational efficiency improvement achieved by designing incident fields with Fourier transform matrix. (a) The probing matrix is created by combining Fourier transform matrix with random phases. (b) In the case that the incident fields are directly generated with random phases, the forward model of scattering is computed by matrix-vector multiplication. Our method designs the incident fields based on Fourier transform matrix. Thus, the forward model of scattering can be computed by FFT.

In Section~\ref{sec:Methods}, we describe the experimental setup for reference-less TM retrieval and introduce the design of probing matrix and the algorithms. In Section~\ref{sec:exp2D}, we demonstrate the experimental result of TM retrieval using intensity images measured at a fixed plane. In Section~\ref{sec:exp3D}, we show the results of the method of TM retrieval with the phase correction by both simulations and experiments. In Section~\ref{sec:discussion} and~\ref{sec:conclusion}, we offer discussion and conclusion remarks.

\clearpage

\section{Methods}
\label{sec:Methods}

\subsection{Experimental setup}

\begin{figure}[ht]
	\centering
	\includegraphics[width=13cm]{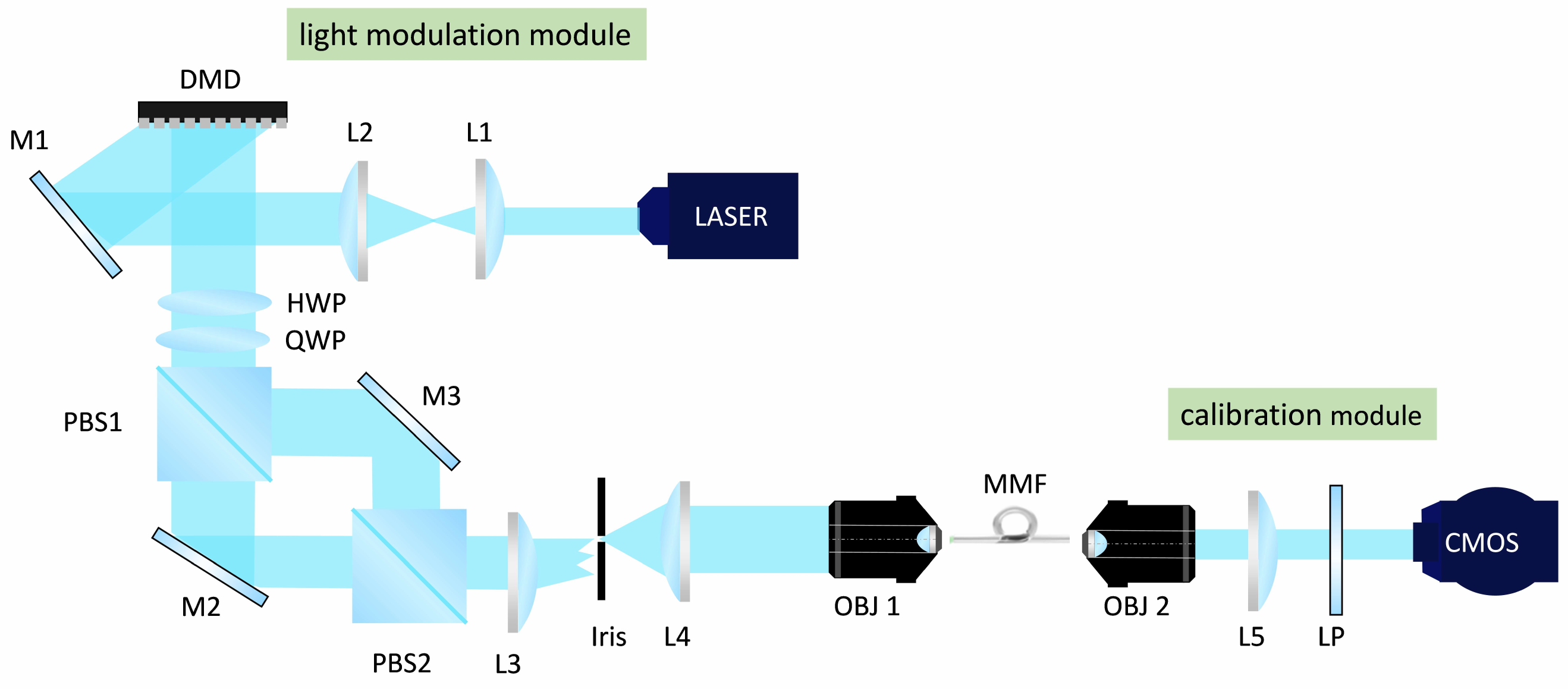} 
	\caption{Experimental setup. The light modulation module on the left of the MMF simultaneously generates the incident complex fields for both polarizations while the calibration module on the right measures the intensity distribution of the transmitted complex fields. The abbreviations are defined as followings: L1-5, lens; DMD, digital micromirror devices; M1-3, mirror; HWP, half-wave plate; QWP, quarter-wave plate; PBS1-2, polarization beamsplitter; OJB1-2, objective lens; LP, linear polarizer.}
	\label{fig:setup}
\end{figure}

The experimental setup is shown in Fig.~\ref{fig:setup}. A MMF of 0.22 numerical aperture (NA) and $50 \ \mu m$ diameter (ChunHui CCS50/125H-F-F-1) is used as the scattering medium. The experimental setup is designed to generate incident complex fields impinging on the proximal end of the MMF and measure the intensity images of the transmitted complex fields at the distal end of the MMF. A collimated laser of 488nm (Precilasers SF-488-0.5-CW) is directed on a DMD (Vialux v9501). It displays the binary hologram obtained by the Lee hologram method~\cite{lee1978computer}. Its entire $1920 \times 1080$ elements provides two $960 \times 960$ regions to modulate both s and p polarizations. A set of half-wave plate and quarter-wave plate interposed between the DMD and the polarization beamsplitter PBS1 turns the light reflected from the DMD into circular polarized light. The s and p lights from PBS1 pass through mirrors M2 and M3 separately, combine by the polarization beamsplitter PBS2, and impinge on the iris after being Fourier transformed by the lens L3. The s and p regions on the binary hologram are programmed with different carrier frequencies, which determines the locations the -1st diffraction order on the Iris. Tuning M2 and M3 shifts the -1st diffraction order of the s and p lights through the pinhole on the Iris. The telescope system formed by the lens L4 and the objective lens OBJ1 (Olympus 10X NA 0.25) focuses the light emitted from the Iris on the distal end of the MMF. Thus, the incident of complex fields with desired phases are generated for both polorizations. The 4f system formed by the objective lens OBJ2 (Olympus 20X NA 0.25) and the lens L5 magnifies the transmitted complex field. A CMOS camera (Basler acA720-520um) captures the intensity image after the light passes through a linear polarizer. The synchronization of the DMD (refresh rate of 16.7 kHz) and the CMOS camera allows the intensity images to be measured at a high framerate, up to 525 frame/s. The objective lens OBJ2 is placed on a piezo stage (Thorlabs CT1P). By moving OBJ2 with the stage, the CMOS camera captures the intensity images at a defocus plane. 

\subsection{TM retrieval}
\label{sec:TMretrieval}
Our method recovers the TM from the intensity images measured from the reference-less experimental setup in Fig.~\ref{fig:setup}. The data-acquisition experiment generates the phase-only incident complex fields by modulating the DMD and measures the intensity images of the transmitted complex fields for TM retrieval. In this section, we define the forward model, formulate the optimization problem of TM retrieval, and develop our proposed TM retrieval method based on FFT.

The intensity images are denoted as $I_{n}(x,y)$, where $n=1,...,N$ and $x,y$ are the spatial coordinates. Each image contains $N_x$ by $N_y$ pixels. The phase modulation is denoted as a vector $e^{j{\theta}_n}$, which has $N_k$  macro-pixels for both polarizations. The forward model of the intensity measurement is written as
\begin{align}
{\bf{I}}_n = {\left| {\bf{TM}} e^{j{\mathbf{\theta}}_n} \right|}^2,
\label{eq:forwardmodel}
\end{align}
where $\bf{I}_n$ is a vector raster-scanned from $I_{n}(x,y)$, $\bf{TM}$ is the transmission matrix, and $\left|\cdot\right|$ takes absolute square of the complex number inside. Note that the transmission matrix has size of $N_x*N_y$ by $N_k$.
%note the size of TM changes after resizing the intensity images.
%{{\bf{I}}_n} = {{\left| {{{\bf{K}}^{H}}{{\bf{H}}_{\bf{n}}}{\bf{b}}} \right|}^2}.

The optimization problem of retrieving TM from the intensity images could be formulated as
\begin{align}
    \min \limits_{\bf{TM}}  \sum \limits_{n} \| {\bf{I}}_n - \left| {\bf{TM}} e^{j{\mathbf{\theta}}_n} \right|^2 \|^2_2,
    \label{eq:optTMlarge}
\end{align}
where $ \|\cdot \|^2_2$ is the squared $L2$ norm of the vector inside. The cost function is the sum of squared error between the intensity measurements and the intensity predicted by the forward model. The optimization problem solves $\bf{TM}$ by minimizing the cost function.

The number of unknown in $\bf{TM}$ is typically very large. It makes the optimization problem in Eq.~\ref{eq:optTMlarge} difficult to solve directly. However, it can be broken down into $N_x*N_y$ smaller optimization problems. Each problem is formulated based on the intensity measurement at one single pixel,
\begin{align}
    \min \limits_{{\bf{tm}}^k}  f({\bf{tm}}^k)=\| {\bf{I}}^k - \left| {\bf{Q}} {\bf{tm}}^k  \right|^2 \|^2_2,
    \label{eq:optTMsmall2}
\end{align}
where
\begin{align}
{\bf{I}}^k= \begin{bmatrix}
{\bf{I}}_1^k\\
{\bf{I}}_2^k\\
\vdots\\
{\bf{I}}_N^k
\end{bmatrix}, \ \
{\bf{Q}}= \begin{bmatrix}
 {e^{j{\mathbf{\theta}}_1}}^T\\
 {e^{j{\mathbf{\theta}}_2}}^T\\
  \vdots\\
 {e^{j{\mathbf{\theta}}_N}}^T
\end{bmatrix}.
\label{eq:optTMsmall2IQ}
\end{align}
Here the column vector ${\bf{tm}}^k$ is the transpose of the $k^{th}$ row of $\bf{TM}$, ${\bf{I}}_n^k$ is the $k^{th}$ element of $\bf{I}_n$, and $T$ denotes transpose.
The vector ${\bf{I}}^k$ contains all of the measurements at the same pixel indexed by $k$ on the intensity image. The matrix $\bf{Q}$ is the so-called probing matrix; each row of $\bf{Q}$ is one of the incident fields. Each optimization problem recovers one row of $\bf{TM}$ from the measurements at the corresponding pixel. Thus, the whole $\bf{TM}$ can be recovered by solving these small optimization problems independently.

Most of the reference-less TM retrieval methods~\cite{huang2021generalizing,n2017controlling,huang2020retrieving} set the phases of the probing matrix $\bf{Q}$ as random numbers. It means the incident fields are modulated with random phases in these methods. By contrast, our method designs the matrix $\bf{Q}$ with Fourier transform matrix,
\begin{align}
{\bf{Q}}= \begin{bmatrix}
{\bf Q}_1\\
 {\bf Q}_2\\
  \vdots\\
 {\bf Q}_M
\end{bmatrix}, 
{\bf Q}_m = {\bf K} \text{diag}(e^{j{\mathbf{\psi}}_m} ).
\label{eq:probingmatrix}
\end{align}
where ${\bf K}$ is the Fourier transform matrix, and $e^{j{\mathbf{\psi}}_m}$ is a $N_k$ by $1$ column vector with its phase set as random numbers. Here $\text{diag}(e^{j{\mathbf{\psi}}_m} )$ is a diagonal matrix whose diagonal entries are from $e^{j{\mathbf{\psi}}_m}$. The phase modulation has $N_{kx} \times N_{ky}$ modes for each polarization, and we have $N_k= 2N_{kx}*N_{ky}$. We set the matrix ${\bf K}$ as the 2D Fourier transform matrix for $N_{kx} \times 2N_{ky}$ matrix. Since ${\bf K}$ is a pure phase matrix, the probing matrix $\bf{Q}$ remains as pure phase. So it can be loaded into the phase modulator to generate desired incident fields, and the total number of measured intensity images is $N=M*N_k$, which increases linearly with $M$. In the case that the phases of the probing matrix $\bf{Q}$ are random, the multiplication of $\bf{Q}$ with a vector has to be computed with matrix-vector multiplication (e.g. Eq.~\eqref{eq:optTMsmall2}). Our method designs the probing matrix $\bf{Q}$ with Fourier transform matrix, the matrix-vector multiplication related to $\bf{Q}$ can be computed with FFT or inverse FFT ( for its complex transpose ${\bf{Q}}^{H}$ ). This advantage can be exploited to accelerate the algorithm for TM retrieval.

We follow the phase retrieval method based on nonlinear optimization~\cite{zhong2016nonlinear} to solve the optimization problem in Eq.~\ref{eq:optTMsmall2}. The optimization is initialized by back propagation,
\begin{align}
{\bf{tm}}^k_0 &=({\bf{Q}}^H{\bf{Q}})^{-1}{\bf{Q}}^H \sqrt{{\bf{I}}^k} \nonumber \\
&=\frac{1}{M}{\bf{Q}}^H \sqrt{{\bf{I}}^k}, \label{eq:backpropagation}
\end{align}
where $\sqrt{{\bf{I}}^k}$ takes element-wise square root of the vector ${\bf{I}}^k$, and $H$ denotes complex transpose. Since  ${\bf K}^H$ is the inverse Fourier transform matrix, the matrix-vector multiplication in ${\bf{Q}}^H \sqrt{{\bf{I}}^k}$ can be computed with FFT.

We derive the first derivative of $f({\bf{tm}}^k)$ with respect to ${\bf{tm}}^k$ as (more details in Appendix~\ref{app:TMretrieval}),
\begin{align}
    {\frac{\partial f} {\partial {\bf tm }^k}}^H = &-4{\bf{Q}}^H\text{diag}({\bf{Q}} {\bf{tm}}^k) ({\bf{I}}^k - \left| {\bf{Q}} {\bf{tm}}^k  \right|^2), \label{eq:derivative} \\
     {\frac{\partial f} {\partial {\bf tm }^k}}^H =& \frac{\partial f}{ \partial {\bf tm }^k_x}^T + j\frac{\partial f}{ \partial {\bf tm }^k_y}^T,
\end{align}
where ${\bf tm }^k_x$ and ${\bf tm }^k_y$ are the real and imaginary parts of ${\bf tm }^k$. The matrix-vector multiplication related to ${\bf{Q}}$ in Eq.~\ref{eq:derivative} can be computed with FFT. 

 The procedure of the algorithm to solve the optimization problem in Eq.~\ref{eq:optTMsmall2} is summarized in Algorithm.~\ref{alg:onerow}. The algorithm has inputs of the intensity measurements at at $k^{th}$ pixel, ${\bf{I}}^k$, and random phase vectors for the probing matrix, $e^{j{\mathbf{\psi}}_m}$. It recovers one row of the TM, ${\bf{tm}}^k$. The estimation is initialized by Eq.~\ref{eq:backpropagation}. After obtaining the error (Eq.~\ref{eq:optTMsmall2}) and gradient (Eq.~\ref{eq:derivative}), the algorithm updates the estimate of ${\bf{tm}}^k$ iteratively by the limited memory Broyden–Fletcher–Goldfarb—Shanno (L-BFGS) method~\cite{wright1999numerical,liu1989limited}. The update iteration stops when a preset maximum iteration number is reached. The matrix-vector multiplication related to ${\bf{Q}}$ in Eq.~\ref{eq:optTMsmall2}, Eq.~\ref{eq:backpropagation} and  Eq.~\ref{eq:derivative} can be efficiently computed with FFT, reducing the computational complexity from $\Theta(N N_k)$ to $\Theta(N\log N_k)$. It also has the benefit of memory efficiency since there is no need to explicitly store the big matrix ${\bf{Q}}$ when solving the inverse problem.

\begin{algorithm}
\caption{Optimization of recovering a row of TM, ${\bf{tm}}^k$.}\label{alg:onerow}
\begin{algorithmic}[1]
\State \textbf{input}: the intensity measurements at $k^{th}$ pixel, ${\bf{I}}^k$, and random phase vectors, $e^{j{\mathbf{\psi}}_m}$, $m=1,...,M$.
\State ${\bf{tm}}^k_0 \gets \text{compute Eq.~\ref{eq:backpropagation} with FFT}$ \Comment{initialization}
\State $iter \gets 0$
\While{$iter < maxiter$}
\State $iter \gets iter + 1$
\State $ f({\bf{tm}}^k_{iter-1}) \gets \text{compute Eq.~\ref{eq:optTMsmall2} with FFT} $
\Comment{error}
\State $ \frac{\partial f} {\partial {\bf tm }^k} |_{{\bf tm }^k_{iter-1}} \gets \text{compute Eq.~\ref{eq:derivative} with FFT} $
\Comment{gradient}
\State $\Delta {\bf{tm}}^k \gets \textit{L-BFGS}[f({\bf{tm}}^k_{iter-1}),\frac{\partial f} {\partial {\bf tm }^k} |_{{\bf tm }^k_{iter-1}}]$
\State ${\bf{tm}}^k_{iter} \gets {\bf{tm}}^k_{iter-1} - \Delta {\bf{tm}}^k$
\EndWhile
\State \textbf{return} ${\bf{tm}}^k_{iter}$
\end{algorithmic}
\end{algorithm}

One may apply the method in Algorithm.~\ref{alg:onerow} on all of $N_x*N_y$ optimization problems in the form of Eq.~\ref{eq:optTMsmall2} and recover the entire TM. However, this could be unnecessary due to the physical properties of the MMF. There is negligible transmitted light on the pixels outside the distal end of the fiber. The complex field at the distal end of the MMF has highest frequency limited by $\text{NA}/ \lambda$, where $NA$ is the numerical aperture of the MMF. Therefore, we design a preprocessing procedure to reduce the number of effective pixels, which in turn brings down the number of optimization problems. 
First, we half-sample the measured intensity images by only keeping the pixels of the odd indexes in the images. Without loss of generality, we assume that the pixel size of the measured intensity images, $ps_{intensity}$, meets the Nyquist sampling theory, 
\begin{align}
ps_{intensity} \leq \lambda/4\text{NA}.
\end{align}
The pixel size of the half-sampled images has $ps_{field}= ps_{intensity}*2$. It meets the sampling requirement of the transmitted complex field, 
\begin{align}
ps_{field} \leq \lambda/2\text{NA}.
\end{align}
And hence, the TM recovered by using the half-sampled intensity images meets the sampling requirement of the transmitted complex field. Compared to the case that the measured intensity images are used directly, this reduces the number of the optimization problems by a factor of 4. Second, from the half-sampled intensity images we obtain a fiber mask which masks out the MMF region. The fiber mask is binary; the pixels within the distal end of the MMF have value of one while the pixels outside of the MMF have value of zero. It is designed in a way that guarantees the one-value region contains 99.9 \% of the total sum of the intensity. For the pixels in the zero-value region, the vectors ${\bf{tm}}^k$ are directly set to zeros, without solving the optimization problem of of Eq.~\ref{eq:optTMsmall2}. At the end, our TM retrieval method only solves the optimization problems for the pixels inside the distal end of the fiber from the half-sampled intensity images. Thus, the number of optimization problems is greatly reduced.
%In case that the pixel size is much smaller than the requirement of Nyquist sampling theory, the images can be rescaled with a larger pixel size which meets the requirement. 

\begin{figure}[htbp]
	\centering
	\includegraphics[width=8.0cm]{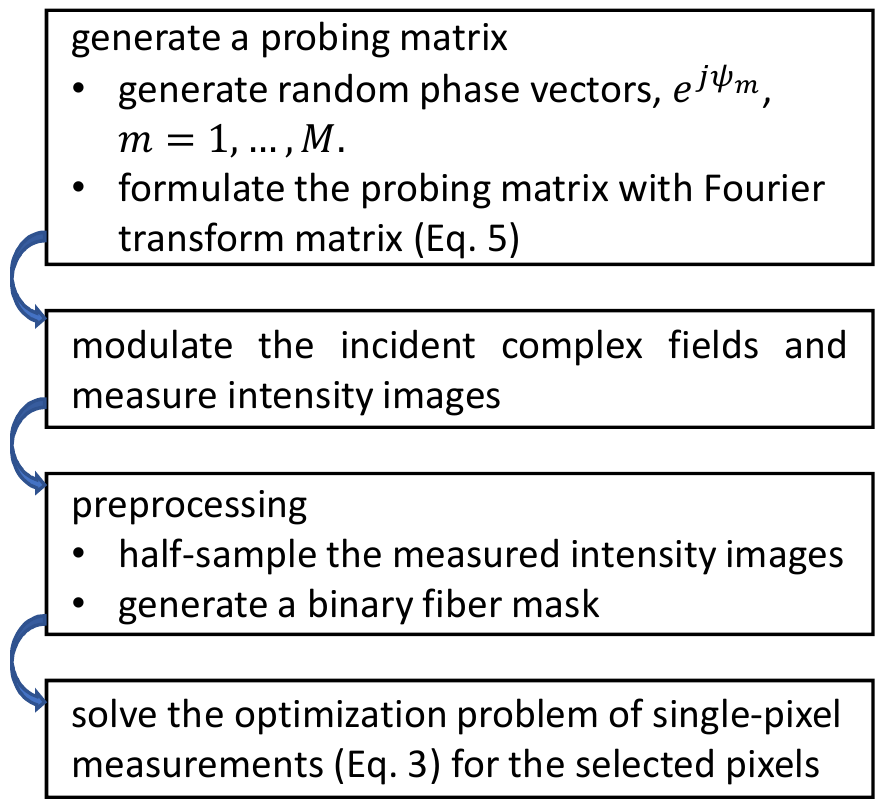} 
	\caption{The full procedure of the TM retrieval method. The intensity images are measured at one fixed camera plane. }
	\label{fig:procedure}
\end{figure}

The full procedure of the TM retrieval method is summarized in Figure.~\ref{fig:procedure}. The optimization problems of Eq.~\ref{eq:optTMsmall2} are independent. So our method solves these optimization problems parallelly with a computer of multiple CPU cores. 

\subsection{Phase correction}
\label{sec:phasecorrection}
The optimization problem in Eq.~\ref{eq:optTMsmall2} has issue of phase ambiguity. It means multiplying an optimal solution of the optimization problem with an arbitrary phase term, $e^{j\phi_0}$ still gives an optimal solution. The phase ambiguity leads to the error of phase offset between the estimated TM and the true TM. It has
\begin{align}
   {\bf{TM}}_{true}= \text{diag}(e^{j\phi}) {\bf{TM}}_{est},
   \label{eq:phaseoffset}
\end{align}
where $ {\bf{TM}}_{true}$ is the true TM, ${\bf{TM}}_{est}$ is the estimated TM, and the vector $e^{j\phi}$ (size $N_x^{half}*N_y^{half}$ by 1) is the phase offset. Note that $N_x^{half}$ and $N_y^{half}$ are the size of the half-sampled intensity image, due to the preprocessing step in Fig.~\ref{fig:procedure}. Although there exists the error of phase offset, ${\bf{TM}}_{est}$ provides sufficient information to generate 2D distributed foci at the plane where the intensity images are measured. However, the error in the phase can cause failure of generating 3D foci at other focal planes.

To solve the issue of phase ambiguity, we propose a phase correction method after ${\bf{TM}}_{est}$ has been obtained by the TM retrieval method in Fig.~\ref{fig:procedure}. Our method corrects the phase offset by using multiple defocus intensity images. After applying random phases to modulate the incident fields, our method measures intensity images at a defocus plane which is $z_d$ away from the distal end of the MMF ($z=0$). The defocus intensity images results from free space propagation of the transmitted complex field at the distal end of the fiber. These defocus intensity images captures the phase information of the transmitted complex fields. Therefore, it is possible to invert the phase offset of the estimated TM from the defocus intensity images.
%In our experimental setup, defocus is achieved by moving the objective lens with a motorized axial stage(???). 

We build the forward model for the inverse problem of the phase offset recovery from the defocus intensity images. The defocus intensity images are denoted with $I_{n}(x,y,z_d)$, where $n=1,...,N_d$. Each intensity image has size of $N_x$ by $N_y$ with pixel size of $ps_{intensity}$. From Eq.~\ref{eq:phaseoffset}, the transmitted complex field at the distal end of the MMF can be expressed as,
\begin{align}
{\bf c}_{n}= \text{diag}({\bf{TM}}_{est} e^{j{\theta}_n^{z_d}}) e^{j\phi},
\end{align}
where $e^{j{\theta}_n^{z_d}}$ is the incident complex field. Note that the transmitted complex field predicted by the estimated TM has size of $N_x^{half}$ by $N_y^{half}$ with pixel size of $ps_{field}$. The vector ${\bf c}_{n}$ is the raster-scanned form of the transmitted complex field. 

The complex field at the defocus plane and the transmitted complex field at the distal end of the MMF are related by defocus propagation. By the theory of angular spectrum propagation~\cite{goodman2005introduction}, the defocus propagation kernel in frequency domain is expressed as,
\begin{align}
h(u,v,z_d)= \exp(j\frac{2\pi}{\lambda} \sqrt{ 1-(\lambda u)^2-(\lambda v)^2 } z_d)p(u,v), \label{eq:angularspectrum}
\end{align}
where $\lambda$ is the wavelength of the illumination, $u$ and $v$ are the spatial frequency coordinates, and $p(u,v)$ is the pupil of the imaging system. The pupil is written as, 
\begin{align}
P(u,v) = \begin{cases}
1 & \quad \lambda \sqrt{u^2+v^2} <= \text{NA}, \\
0 & \quad  \lambda \sqrt{u^2+v^2} > \text{NA}.
\end{cases}
\end{align}

Next we obtain the vectors ${\bf{I}}_n^{z_d}$, ${\bf h}$ which are raster-scanned from $I_{n}(x,y,z_d)$ and $h(u,v,z_d)$. The forward model of the defocus intensity can be expressed as
\begin{align}
 {\bf{I}}_n^{z_d}= \left| {\bf K}_2^H {\bf P}\text{diag}({\bf h}){\bf K}_1 \text{diag}({\bf{TM}}_{est} e^{j{\theta}_n^{z_d}}) e^{j\phi} \right|^2,
 \label{eq:forwarddefocusImage}
\end{align}
where ${\bf K}_1$ is the Fourier transform matrix for $N_x^{half} \times N_y^{half}$ matrix, and ${\bf P}$ is for zero padding in the Frequency domain, ${\bf K}_2^H$ is the inverse Fourier transform matrix for $N_x \times N_y$ matrix. The measured intensity has size of $N_x$ by $N_y$ while the transmitted complex field has size of $N_x^{half}$ by $N_y^{half}$. The zero padding here adds zeros in frequency domain which results in doubling the number of pixels in both dimensions. It has effect of reversing the half-sample step in Fig.~\ref{fig:procedure}.

The optimization problem of solving the phase offset from the defocus intensity images is formulated as,
\begin{align}
   \min \limits_{\phi}  g(\phi)=\sum \limits_{n} \|  {\bf{I}}_n^{z_d}- \left|   {{\bf A}_n} e^{j\phi} \right|^2 \|^2_2,
    \label{eq:optphase}
\end{align}
where $ {{\bf A}_n}= {\bf K}_2^H {\bf P}\text{diag}({\bf h}){\bf K}_1 \text{diag}({\bf c}_{n}) $ . The cost function is defined as the squared error between the measured defocus intensity and the intensity predicted with the phase offset. 

The first derivative of the optimization problem in Eq.~\ref{eq:optphase} is written as,
\begin{align}
{\frac{\partial g} {\partial \phi}}^H =  \sum \limits_{n} \text{real}(-4 \text{diag}(-je^{-j\phi}) {{\bf A}_n}^H \text{diag}({{\bf A}_n}e^{j\phi})(  {\bf{I}}_n^{z_d} - \left| {{\bf A}_n} e^{j\phi} \right|^2)).
\label{eq:phasederive}
\end{align}
More details can be found in Appendix~\ref{app:Phasecorrection}. The matrix-vector multiplication related to ${{\bf A}_n}$ in Eq.~\ref{eq:optphase} and Eq.~\ref{eq:phasederive} can be computed with FFT, without explicitly forming the big matrices. With th cost function in Eq.~\ref{eq:optphase} and the first derivative in Eq.~\ref{eq:phasederive}, our method uses the L-BFGS method~\cite{wright1999numerical,liu1989limited} to recover the phase offset from the defocus intensity images. 

%The procedure to recover the phase offset is summarized in Algorithm.~\ref{alg:phase}.

\iffalse
\begin{algorithm}
\caption{phase correction from defocus intensity images.}\label{alg:phase}
\begin{algorithmic}[1]
\State \textbf{input}: the defcous intensity images (${\bf{I}}_n^{z_d}$), the estimated TM by the algorithm in Fig.~\ref{fig:procedure} (${\bf{TM}}_{est}$), and the incident complex fields ($e^{j{\theta}_n^{z_d}$).
\State $\phi \gets 0\text{s} $ \Comment{initialization}
\State $iter \gets 0$
\While{$iter < maxiter$}
\State $iter \gets iter + 1$
\State $ f({\bf{tm}}^k_{iter-1}) \gets \text{compute Eq.~\ref{eq:optTMsmall2} with FFT} $
\Comment{error}
\State $ \frac{\partial f} {\partial {\bf tm }^k} |_{{\bf tm }^k_{iter-1}} \gets \text{compute Eq.~\ref{eq:derivative} with FFT} $
\Comment{gradient}
\State $\Delta {\bf{tm}}^k \gets \textit{L-BFGS}[f({\bf{tm}}^k_{iter-1}),\frac{\partial f} {\partial {\bf tm }^k} |_{{\bf tm }^k_{iter-1}}]$
\State ${\bf{tm}}^k_{iter} \gets {\bf{tm}}^k_{iter-1} - \Delta {\bf{tm}}^k$
\EndWhile
\State \textbf{return} ${\bf{tm}}^k_{iter}$
\end{algorithmic}
\end{algorithm}
\fi

%while the calibration module on the lelf measures the transmitted intensity

\section{Results for TM retrieval using intensity images at one measurement plane}
\label{sec:exp2D}
%\subsection{Simulations}
In this section, we verify the TM retrieval algorithm in Fig.~\ref{fig:procedure} by both simulations and experiments. In the simulation, a TM of size 9216 by 8192 was used to generate simulated data. The TM had been measured experimentally by the method of off-axis holography~\cite{vcivzmar2011shaping}, for a MMF of 0.22 NA and 50 $\mu m$ diameter. The off-axis holography method uses an external reference beam to measure the transmitted complex fields. The incident complex fields had $64\times64$ phase modulation modes for each polarization. The transmitted complex fields at the distal end were sampled with $96\times96$ pixels. We simulated 7 data sets with $M=3,4,5,6,7,8,9$, where $M$ represents the total number of $\bf{Q}_m$ used (Eq.~\ref{eq:probingmatrix}). The matrix $\bf{K}$ in the probing matrix (Eq.~\ref{eq:probingmatrix}) was set as the Fourier transform matrix for $64 \times 128$ matrix. The matrix-vector multiplication $\bf{Q}{\bf{tm}}^k$ in Eq.~\ref{eq:optTMsmall2} can be computed with FFT. Therefore, we used $\left|\bf{Q}{\bf{tm}}^k\right|^2$ to generate the simulated data, instead of the forward model in Eq.~\ref{eq:forwardmodel}. The total number of measured intensity images of a simulated data set is $M*8192$, which increases linearly with $M$. 

\begin{figure}[htbp]
	\centering
	\includegraphics[width=\linewidth]{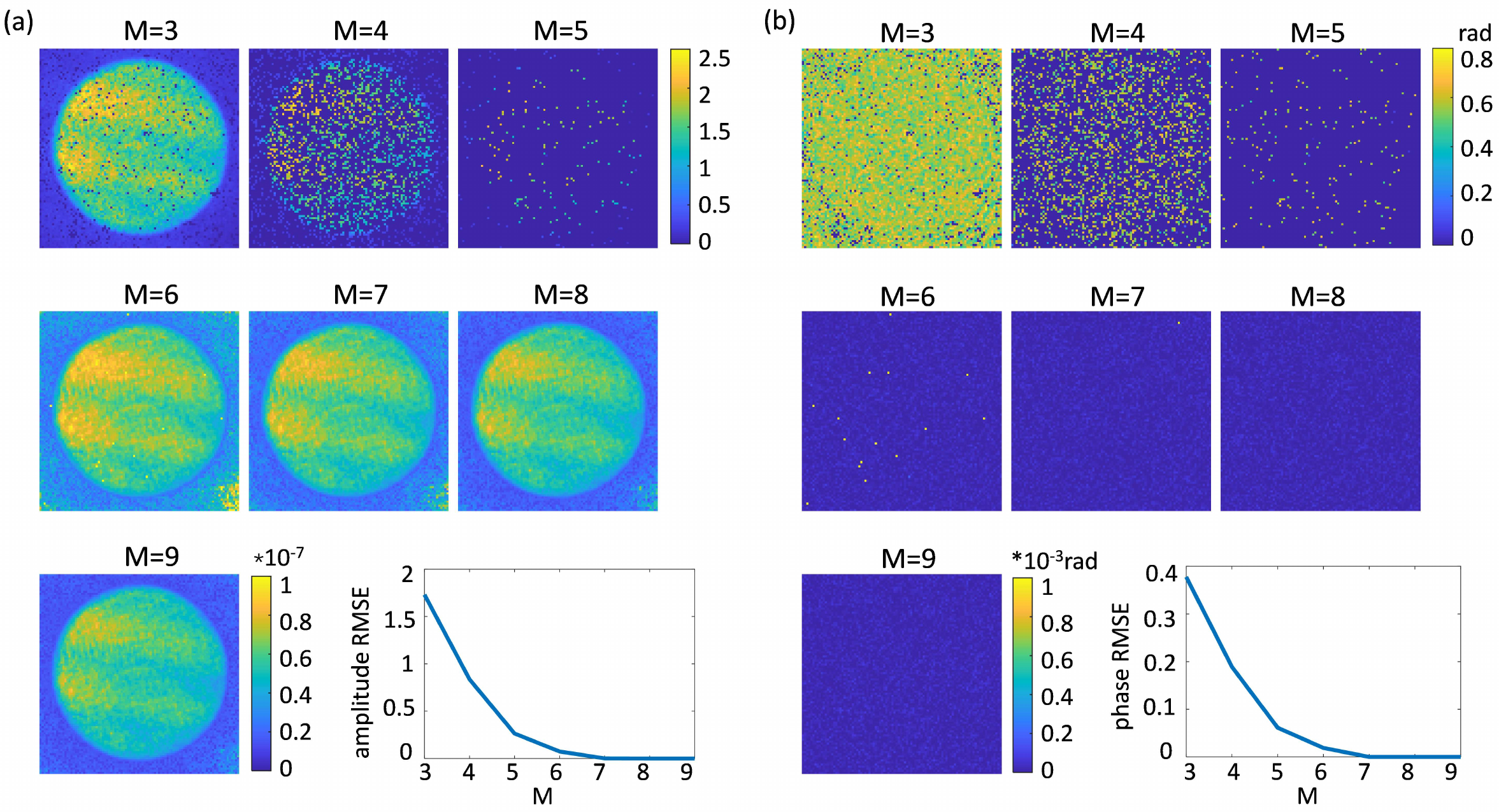} 
	\caption{Error of recovered TM using the simulated data sets. (a) Normalized amplitude error of the recovered TM for data sets of different M. The errors of $M=3,4,5$ shares the same color bar on the top right while the errors of other data sets share the color bar on the bottom. The plot at the bottom right shows the RMSE of amplitude of the recovered TM. (b) Phase error of the recovered TM for data sets of different M. The errors of $M=3,4,5$ shares the top right color bar while the errors of the other data sets share the bottom color bar. The plot shows the RMSE of phase of the recovered TM for different M. }
	\label{fig:2Dsim}
\end{figure}

We ran the TM retrieval algorithm on each of the simulated data sets. We recovered the entire TM by applying the method in Algorithm.~\ref{alg:onerow} on all of the 9216 pixels, without the preprocessing step in Fig.~\ref{fig:procedure}. For each data set, the optimization problems were solved in a parallel manner on a computer with 32 CPU cores (Intel Xeon Gold 5218 2.3GHz). For the data set of M=8, it takes 376.4s to retrieve the entire TM of size 9216 by 8192. Figure.~\ref{fig:2Dsim} compares the error of the recovered TM using the data sets of different measurement sizes. It shows the root mean square error (RMSE) of both amplitude (Fig.~\ref{fig:2Dsim}(a)) and phase (Fig.~\ref{fig:2Dsim}(b)) of the recovered TM. Each row of the recovered TM is compared with its true value, and the errors of all of the 9216 rows are organized in $96\times96$ grids which are shown in Fig.~\ref{fig:2Dsim}. The phase error is obtained by subtracting phase of each row of the recovered TM with the true values after removing the constant phase offset (Eq.~\ref{eq:phaseoffset}). For $M=3$, most rows of the recovered TM have large errors. For $M=4,5,6$, a few of the rows of the recovered TM have large errors; there are random bright spots (meaning large errors) in the images at the top row of Fig.~\ref{fig:2Dsim}(a) and (b). However, these speckles disappear as M increases. The error of the recovered TM becomes negligibly small for M= 7, 8, 9. The bottom right images on Fig.~\ref{fig:2Dsim}(a) and (b) shows the RMSE of both amplitude and phase of the recovered TM converge to zero for M= 7, 8, 9. The simulation demonstrates  the proposed TM retrieval method is able to efficiently recover the TM from the intensity images measured at one imaging plane with negligible errors.

Table~\ref{tab:time} shows the improvement of computational time by the proposed TM retrieval with FFT. The central $32\times 32$ pixels of the $96\times 96$ pixels of the data set of M=8 was used to access the computational time of the TM retrieval algorithms. The TM retrieval algorithm without FFT replaces the FFT in Algorithm.~\ref{alg:onerow} with matrix-vector multiplication. The TM retrieval algorithm without FFT recovers the $1024\times 8192$ TM with 43664.1 seconds (12.1 hours). However, the proposed TM retrieval algorithm implemented with FFT recovers the same-size TM with 35.4 seconds. For the proposed algorithm, each row of the TM takes 0.035s on average. By using FFT, the proposed TM retrieval algorithm achieves 1200x speed-up.

\begin{table}[htbp]
\centering
\caption{\bf The TM retrieval algorithm with FFT achieves 1200x speed-up.}
\begin{tabular}{ccc}
\hline
Methods & $1024 \times 8192$ TM/s & average/s \\
\hline
TM retrieval without FFT & 43664.1 & 42.641\\
TM retrieval with FFT& 35.4 & 0.035 \\
\hline
\end{tabular}
  \label{tab:time}
\end{table}

\begin{figure}[htbp]
	\centering
	\includegraphics[width=\linewidth]{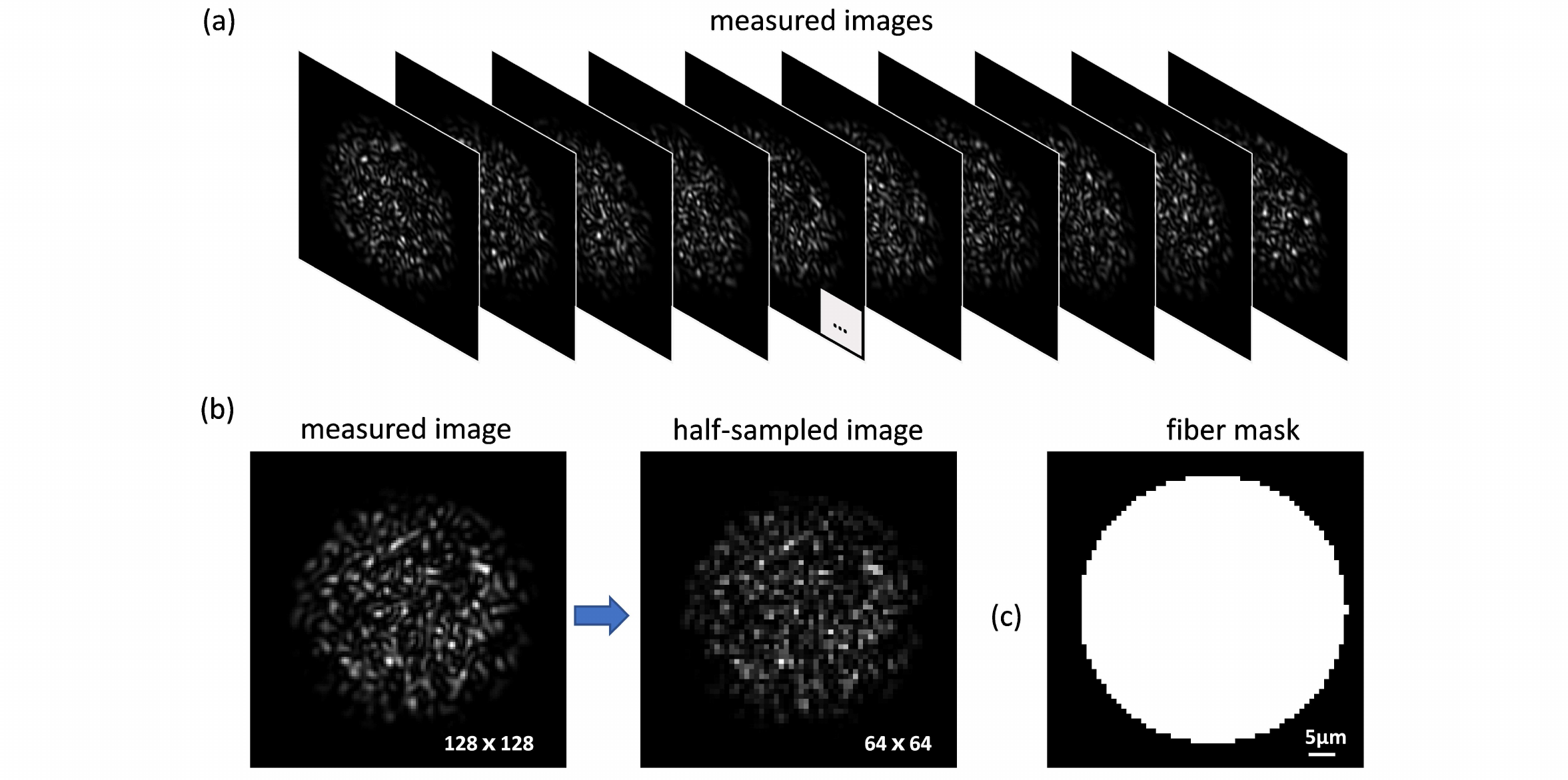} 
	\caption{Measured images and the preprocessing step. We give an example of the measured images and the preprocessing step using the data set of $M=7$. (a) A series of measured speckle intensity images. (b) The prepossessing step half-samples the measured $128 \times 128$ images into $64 \times 64$ images. (c) The binary fiber mask . The white region of the mask indicates the distal end of the MMF fiber. }
	\label{fig:2Dexp1}
\end{figure}

%\subsection{Experiments}
In the experiment, we used a MMF of 0.22 NA and 50 $\mu m$ diameter. The illumination was laser of 488 nm. The DMD achieved $64 \times 64$ phase modulation for each polarization, resulting 8192 modes in total. The cameras measured the intensity images at the distal end of the MMF. We test the TM retrieval algorithm for the cases of $M=3,4,5,6,7,8$. Each case followed the procedure in Fig.~\ref{fig:procedure} to recover the TM. For each case, we generated the probing matrix with random phase vectors and Fourier transform matrix by Eq.~\ref{eq:probingmatrix}. The matrix $\bf{K}$ was set as the Fourier transform matrix for $ 64 \times 128$ matrix. The phase of the probing matrix was loaded into the DMD, and a series of $M*8192$ images (Fig.~\ref{fig:2Dexp1}(a)) were measured by the CMOS camera. Each image has $128 \times 128$ pixels with pixel size of 0.47 $\mu m$. The preprocessing step half-sampled the measured images and obtained images of $64 \times 64$ (Fig.~\ref{fig:2Dexp1}(b)). From the preprocessed images, we calculated the fiber mask (Fig.~\ref{fig:2Dexp1}(c)) which covers 99.9 \% of the total energy. The white region of the mask covers the distal end of the MMF fiber. Only for the pixels inside the fiber mask, the TM were retrieved by Algorithm.~\ref{alg:onerow} from the preprocessed images of each data set. 

Here we give an example of the case of M=7. The probing matrix has size of $57344 \times 8192$. The preprocessed intensity images contains 57344 images of size $64 \times 64$. The number of pixels inside the white region of the fiber mask is 2286, so the retrieved TM has size of $2286 \times 8192$. For each pixel inside the fiber mask, an optimization problem in form of Eq.~\ref{eq:optTMsmall2} is formulated; it has inputs of the intensity measurement at the corresponding pixel (a vector of 57344 by 1) and the random phase vectors used to generate the probing matrix. All these 2286 optimization problem were solved parallelly on the computer with 32 CPU cores. For the cases of M=7, the computer takes 112.9 seconds to solve the optimization problems in TM retrieval.

\begin{figure}[htbp]
	\centering
	\includegraphics[width=\linewidth]{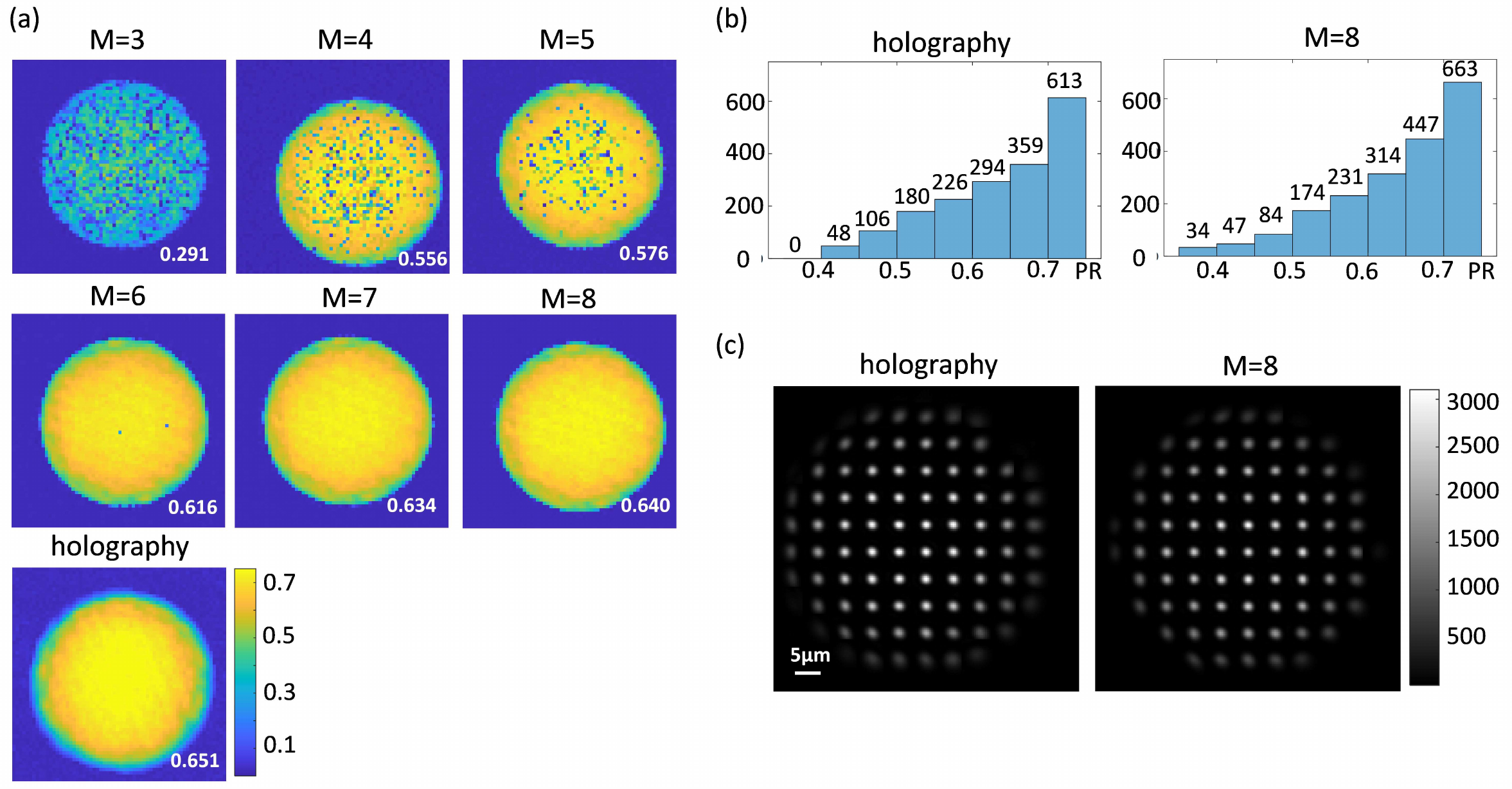} 
	\caption{Comparison of the foci generated by using the recovered TM and the TM measured by the off-axis holography method. (a) The recovered PR of $M=3,4,5,6,7,8$ and the recovered PR of the holography method. For the cases of $M=7$,and $8$, the PR of the foci is near to that of the case of holography. The number inside the image is the average of the top 2000 PR. (b) Histogram of the top 2000 PR. The TM of $M=8$ has 1424 foci which have PR higher than 0.60 while the holgoraphy method has 1266 foci above 0.60. (c) The sum projection of selected foci. }
	\label{fig:2Dexp2}
\end{figure}

The accuracy of the recovered TM was test by the ability to generate foci at the measurement plane. After the TM was retrieved for each case, we uploaded the phase of the conjugate complex of the recovered TM into the phase modulator and sequentially modulated the incident field with the phase row by row to generate intensity images at the camera. When the displayed phase of a row of the retrieved TM matches with the true TM of the imaging system, a foci is generated at the camera. In order to evaluate the quality of the foci, we measured two images ($128 \times 128$ pixels) for each foci with exposure time of 70 $\mu s$ and 1400 $\mu s$, and calculate the power ratio (PR) of the foci by combining these two images. The PR is the ratio of the signal to the total energy. The signal is the sum of  the $7 \times 7$ pixels near the peak of the foci by using the 70 $\mu s$ image, while the total energy is the sum of the signal and the background (outside the $7 \times 7$ pixels), which is calculated by using the 1400 $\mu s$ image and scaled by 20. The PR reflects the quality of the foci, and hence experimentally shows the correctness of the retrieved TM. We measured a TM by the off-axis holography method with an external reference beam and acquired the corresponding foci images. The result by the holography method acts as a reference for our method. Figure~\ref{fig:2Dexp2}(a) shows the PR of the foci of cases of different $M$ and the holography method. For the cases $M=4,5,6$, there are several foci which have low PR. However, for the cases of $M=7,8$, the overall quality of the foci is near to that of the holography method. The average PR of the case of $M=8$ is 0.64, which is slighter smaller than that of the case of holography (0.651). Figure~\ref{fig:2Dexp2}(b) and (c) further compares the cases of $M=8$ and holography by showing the distribution of the power ratio and a sum projection of several selected foci. The TM retrieval method by $M=8$ have more foci of PR above 0.60 than that of the holography method. And hence, the accuracy of the TM recovered by our proposed reference-less method is validated by comparing with the holography method.

\section{Results for TM retrieval with phase correction}
\label{sec:exp3D}

\begin{figure}[hbtp]
	\centering
	\includegraphics[width=\linewidth]{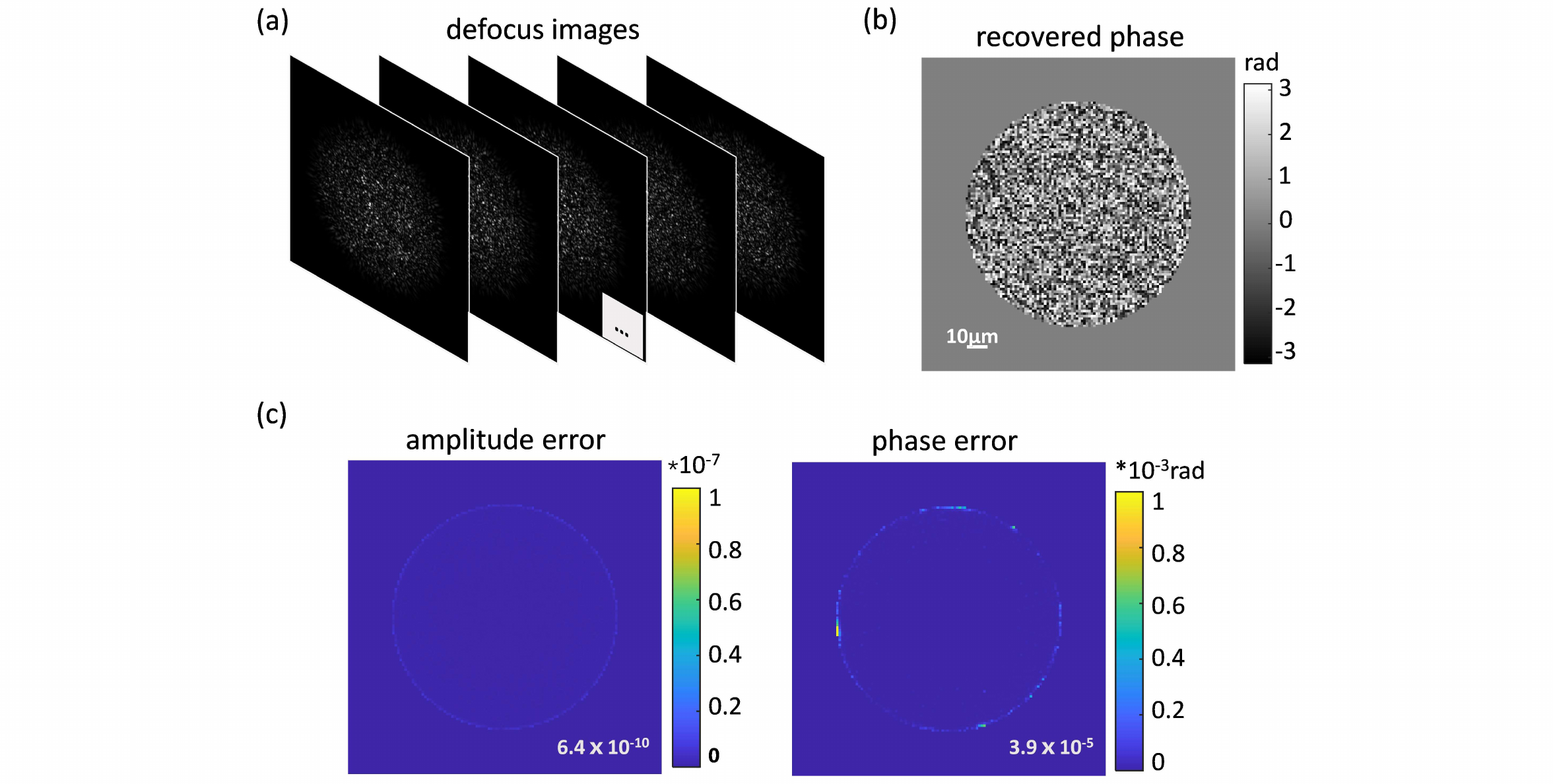} 
	\caption{Simulation for the TM retrieval algorithm with phase correction. (a) Defocus intensity images measured for the phase correction. (b) Recovered phase offset by the phase correction algorithm. (c) Amplitude error and phase error of the recovered TM with phase correction. The amplitude error is obtained by subtracting the amplitudes of the corrected TM with the true TM. The phase error is the difference between the phases of the corrected TM and the true TM after removing a constant phase offset. The RMSE of all rows of the TM are organized in 128 by 128 grids, corresponding to the distal end of the MMF. The numbers inside the images are the RMSE over all rows. }
	\label{fig:3Dsim}
\end{figure}

In this section,  we validate the TM retrieval algorithm with phase correction (Section~\ref{sec:phasecorrection}). In the simulation, we used a simulated TM of size $16384 \times 9216$ for a MMF with 0.22 NA and 100 $\mu m$ diameter, generated by solving Maxwell's Equations. The transmitted complex fields of the MMF are sampled by $128\times 128$ grids with pixel size of 1.1667 $\mu m$, and the wavelength of illumination is 532 nm. We designed a probing matrix with a 2D Fourier transform matrix for $96 \times 96$ matrix and M=9. A series of 82944 images of size $128 \times 128$ were generated at the distal end of the fiber $z= 0 \ \mu m$. Then we simulated 50 defocus images (Fig.~\ref{fig:3Dsim}(a)) at $z_d= 50 \ \mu m$ away from the distal end of the MMF. Each image has $256 \times 256$ pixels with pixel size of 0.5833 $\mu m$. The incident complex fields were obtained by 50 random phases, and the defocus intensity images were generated by Eq.~\ref{eq:forwarddefocusImage}.

First, we followed the preprocessing step and the optimization step of the procedure in Fig.~\ref{fig:procedure} to recover the TM. In the preprocessing step, the half-sample step was not performed since the pixel size already meets the sampling requirement of the complex field. A fiber mask was generated, resulting 6668 selected pixels inside the white region. For the selected pixels, the optimization problems in the form of Eq.~\ref{eq:optTMsmall2} were solved, and the rows of the recovered TM corresponding to the black region in the mask were set to zeros. Thus, a recovered TM was obtained but has the error of phase offset, since the measured intensity images were at one fixed plane. Next, the phase offset were solved from the defocus images and the recovered TM by the phase correction algorithm in Section.~\ref{sec:phasecorrection}. The computational time for the TM retrieval and the phase correction were 332.9 seconds and 85.2 seconds, respectively. Figure \ref{fig:3Dsim}(b) shows the recovered phase by the algorithm. Finally, we compensated the phase offset error of the recovered TM by using the recovered phase. The amplitude and phase RMSE of the recovered TM with phase correction is $6.4 \times 10^{-10}$ and $3.9 \times 10^{-5}$, respectively. The error between the recovered TM with phase correction and the true TM is small as shown in Fig.~\ref{fig:3Dsim}(c). 

\begin{figure}[tb]
	\centering
	\includegraphics[width=\linewidth]{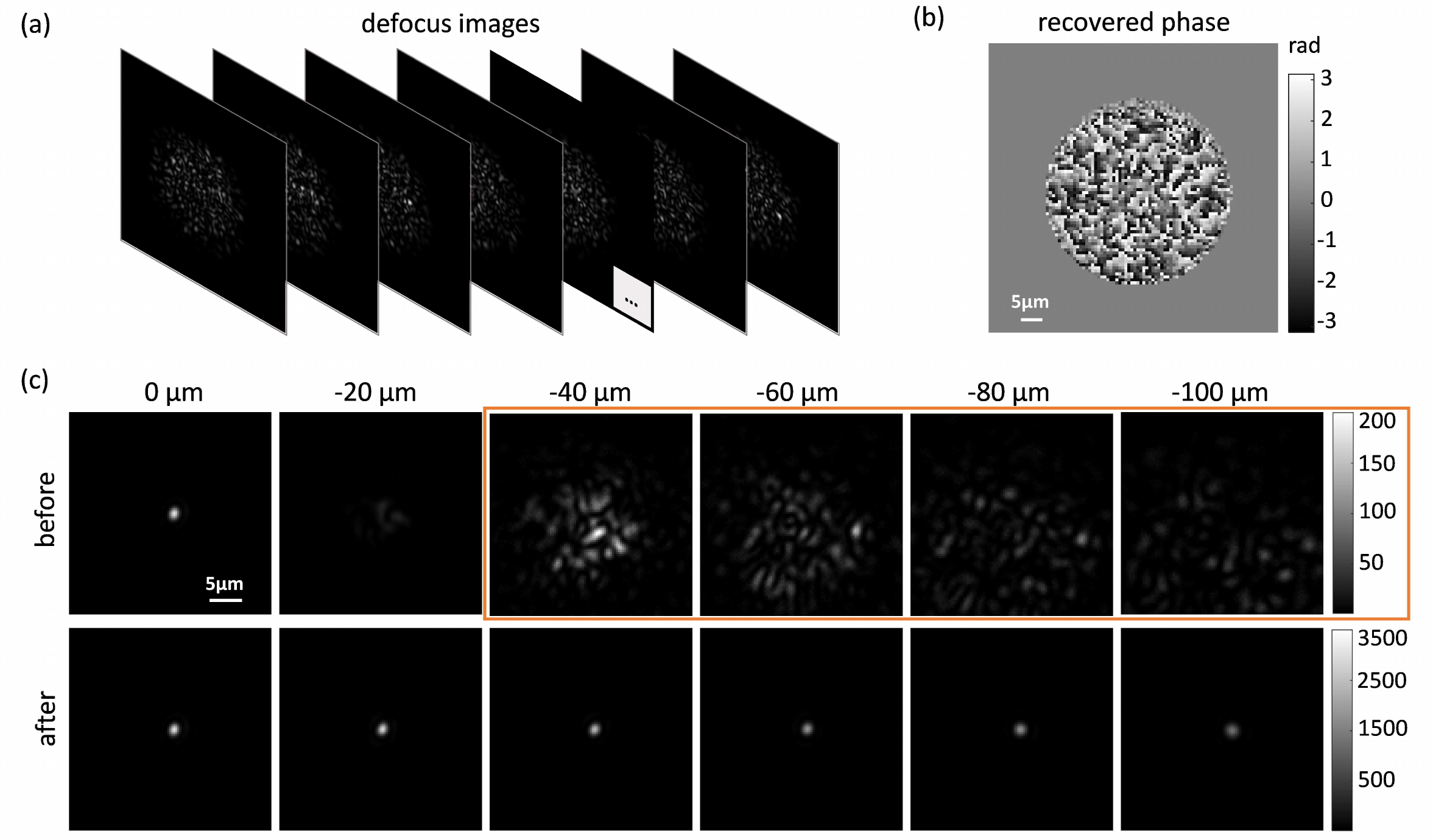} 
	\caption{Correction of the phase offset error in the TM by using defocus intensity images. (a) A stack of defocus images. (b) Recovered phase offset by the phase correction algorithm. (c) The intensity images generated using the TM with the error of phase offset and the recovered TM with phase correction. The top row shows the measured intensity images using the TM with the error of phase offset. The foci scattered at large defocus distances. The bottom row shows the measured intensity images using the recovered TM with phase correction. The images inside the orange box shares the same color bar at top right while the other images share the color bar at bottom right.}
	\label{fig:3Dexp1}
\end{figure}

We further validated the TM algorithm with phase correction by experimentally displaying 3D foci. In the experiment, we used a MMF of 50 $\mu m$ diameter and 0.22 NA, and illumination wavelength of 488 nm. The phase modulation on DMD had $64 \times 64$ modes for each polarization. We designed a probing matrix using Fourier transform matrix for $64 \times 128$ matrix, and M=8. After modulating the DMD with the phase of the probing matrix, we sequentially measured 65536 intensity images at the distal end of the MMF (z=0 $\mu m$). Each image has $192 \times 192$ pixels with pixel size of 0.4182 $\mu m$. In order to correct the phase offset, we measured 50 images of $192 \times 192$ at 40 $\mu m$ away from the distal end (Fig.~\ref{fig:3Dexp1}(a)). The defocus images were measured after applying 50 random phases on the phase modulator.

We first recovered a TM from the intensity images measured at z=0 $\mu m$ by the proposed method in Fig.~\ref{fig:procedure}. In the preprocessing step, we half-sampled the images to size of 96 by 96, and generated a fiber mask which has 3015 pixels inside the white region of the mask. By solving the optimization problems, the TM retrieval algorithm obtained a TM. It has the error of the phase offset since the intensity images were measured at one fixed plane. Next, the algorithm of phase correction recovered the phase offset (Figure.~\ref{fig:3Dexp1}(b)) from the defocus intensity images. The recovered phase offset was used to correct the error of phase offset in the recovered TM. The computational times for the algorithm of TM retrieval and the algorithm of phase correction are 199.3 and 20.6 seconds, respectively.

\begin{figure}[ht]
	\centering
	\includegraphics[width=\linewidth]{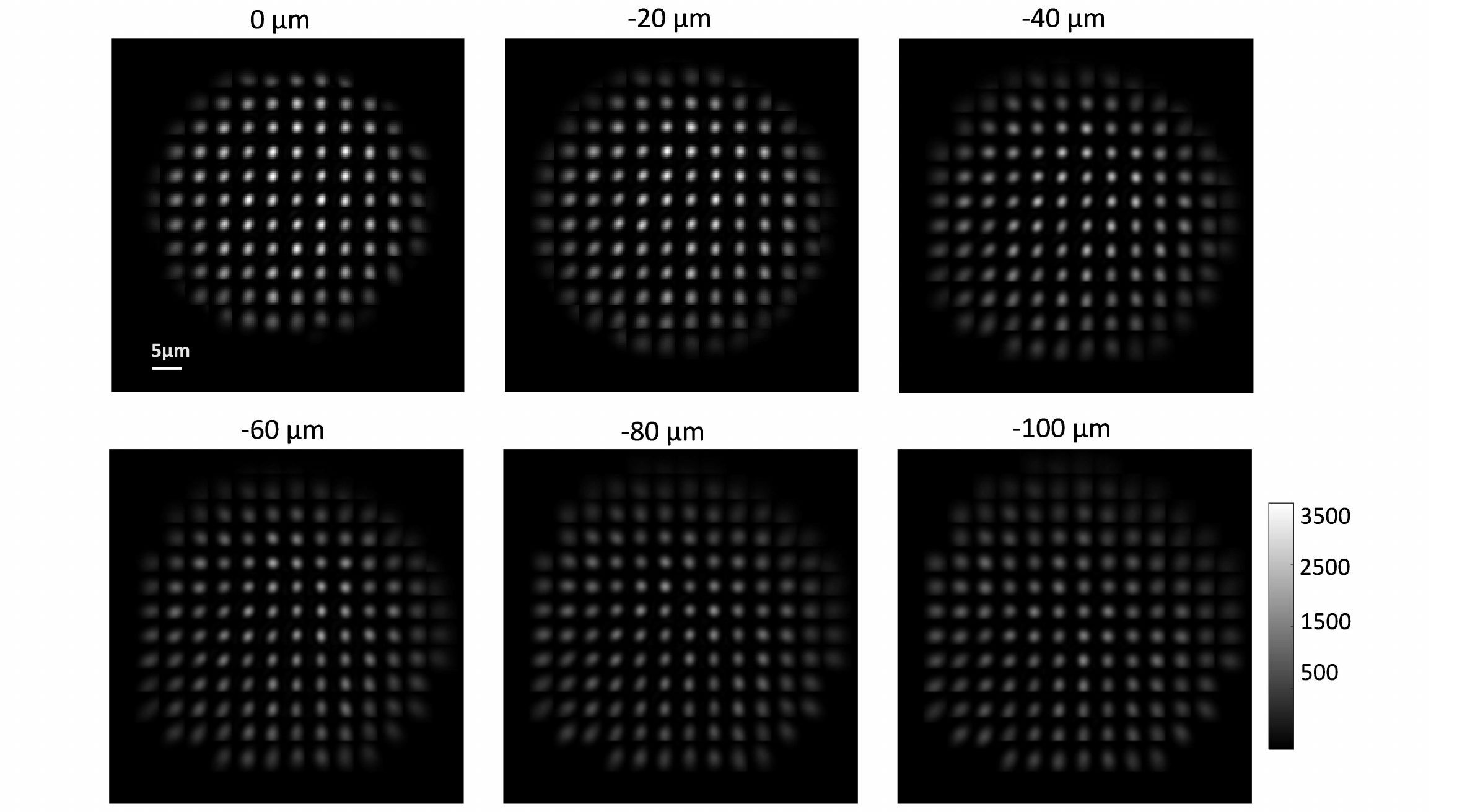} 
	\caption{Sum projection of selected foci measured at different defocus distances. }
	\label{fig:3Dexp2}
\end{figure}

We test the recovered TM by generating 3D foci on the imaging system. The propagated TM at a defocus distance could be obtained by adding the recovered TM at $z=0$ with a free space defocus propagation. We generated the two sets of propagated TM at $z= 0,-20,-40,-60,-80,-100 \ \mu m$, by using the recovered TM with the error of phase offset and the recovered TM with phase correction. We sequentially applied the phases of complex conjugate of the propagated TM to the DMD, and measured intensity images at the corresponding defocus distances. Figure.~\ref{fig:3Dexp1}(c) compares the intensity images measured at different defocus distances for the foci at the center of the images. For the case of the TM with phase error, the foci could be observed at the image center for $z= 0 \ \mu m$, but it quickly scattered into random patterns in the images measured at other defocus distances (top row of Fig.~\ref{fig:3Dexp1}(c)). The phase offset error causes the failure in generating 3D foci. By contrast, the propagated TM generated by using the recovered TM with phase correction successfully generate the foci at defocus distances (bottom row of Fig.~\ref{fig:3Dexp1}(c)). As the defocus distances increase from $0 \ \mu m$ to $100 \ \mu m$, the PR of the foci reduces from 0.60 to 0.51. The decrease of foci brightness could be caused by the defocus propagation. It adds more correlation for the rows of propagated TM corresponding to the neighborhood pixels. Figure.~\ref{fig:3Dexp2} shows the sum projection of selected foci at different defocus distances generated by using the recovered TM with phase correction. This validates the accuracy of the recovered TM with phase correction.

\section{Discussion}
~\label{sec:discussion}
The optimization problem in Eq.~\ref{eq:optTMsmall2} is a phase retrieval problem. The cost function of the phase retrieval problem is formulated based on intensity difference, which is suitable for the assumption that the intensity measurements are polluted by Gaussian noise. With the assumption of Poisson noise, the cost function can be formulated with amplitude difference~\cite{yeh2015experimental}. Many algorithms have been proposed for the phase retrieval problem, including gradient descent~\cite{fienup1982phase}, Gerchberg-Saxton~\cite{gerchberg1972practical}, Kalman filtering~\cite{waller2011phase}, L-BFGS~\cite{brady2006nonlinear,zhong2016nonlinear}, modified Gauss Newton~\cite{zhong2016nonlinear}, Wirtinger flow~\cite{candes2015phase}, prVBEM~\cite{dremeau2015phase}, PhaseLift~\cite{candes2013phaselift}, reweighted amlitude flow~\cite{wang2018phase}, PhaseMax~\cite{goldstein2018phasemax}. The L-BFGS method is a second order optimization method which was shown to converge faster than the first order methods such as gradient descent or Gerchberg-Saxton in phase retrieval from defocus images~\cite{zhong2016nonlinear} and Fourier ptychography~\cite{yeh2015experimental}. In this work, we used the intensity-based cost function and the L-BFGS method. A fair assessment of the formulation of the cost function and the optimal choice of the algorithm for the phase retrieval problem in the TM retrieval is out of the scope of this work.

This work proposed to design the probing matrix $\bf{Q}$ ($N \times N_k $) with Fourier transform matrix. By using FFT, the computational complexity of the matrix-vector multiplication related to $\bf{Q}$ and $\bf{Q}^H$ reduces from $\Theta(N N_k)$ to $\Theta(N \log(N_k))$. Here we give an example of the number of modulation modes $N_k=8192$ and the number of measurement $N=65536$. The matrix $\bf{Q}$ has size of $65536 \times 8192$. The computational complexity reduces from  $\Theta(65536 \times 8192)$ to $\Theta(65536 \times 13)$, and it is memory-efficient without storing  $\bf{Q}$. The computation related to the probing matrix $\bf{Q}$ is mostly inevitable in the algorithms of the phase retrieval problem. For example, gradient descent based algorithms have to compute the cost function and gradient descent. The computational complexity of these algorithms is lower-bounded by $\Theta(N N_k)$, due the matrix-vector multiplication related to $\bf{Q}$. It is higher than that of our proposed method using FFT, $\Theta(N \log(N_k))$. However, applying the similar FFT-based scheme in these algorithms could further reduce the computational complexity.

\section{Conclusion}
\label{sec:conclusion}
We have demonstrated a novel method for reference-less TM retrieval and validated the method by both simulations and experiments. We proposed to design the probing matrix based on Fourier transform matrix and developed an efficient TM retrieval algorithm based on FFT. We demonstrated that the proposed method can recover the TM of size $2286 \times 8192$ with 124.9 seconds for the MMF of 0.22 NA and $50 \ \mu m$ diameter by the computer of 32 CPU cores. We also proposed the algorithm which corrects the error of phase offset in the TM retrieval by using the defocus intensity images. We validated the phase correction algorithm by generating 3D foci with the experimental setup.

 With the advantage of computational efficiency and the correction of phase offset, we envision our method can be used in a broad range of TM retrieval related applications. Our method is suitable for the case where the interferometric setup is difficult to build. For example, one can use our method to calibrate the TM of a long optical fiber in optical communication, for which a external reference beam is hard to obtain. One can also use our method to simplify the experimental setup by removing the reference beam and achieve 3D volumetric imaging through endoscopy based on MMF. We demonstrated our method with a computer of 32 CPU cores. A computer with more parallel cores can further reduce the computational time. Although we verified our method by using MMF, one may adapt our method to measure the TM of other scattering media or imaging systems.

\begin{appendices}

\setcounter{table}{0} % 将表格计数设置为0
\setcounter{figure}{0} % 将图片计数设置为0
\setcounter{equation}{0} % 将公式计数设置为0
\renewcommand{\thetable}{\thesection\arabic{table}} %在表格序号前面加上章节号，如：A-1，B-1 
\renewcommand{\theequation}{\thesection\arabic{equation}} %在公式序号前加上章节号，如：A-1

%The appendix section provides the deviation of the first derivative in the TM retrieval algorithm and the phase correction algorithm.

\section{Derivation of the first derivative in the TM retrieval}
\label{app:TMretrieval}
\setcounter{table}{0} % 将表格计数设置为0
\setcounter{figure}{0} % 将图片计数设置为0
\setcounter{equation}{0} % 将公式计数设置为0

The optimization problem to recover one row of TM is expressed as,
\begin{align}
    \min \limits_{{\bf{tm}}^k}  f({\bf{tm}}^k)=\| {\bf{I}}^k - \left| {\bf{Q}} {\bf{tm}}^k  \right|^2 \|^2_2.
    \label{eq:apoptTMsmall2}
\end{align}
Next, we define,
\begin{align}
   {\bf F}= {\bf{I}}^k - \left| {\bf{Q}} {\bf{tm}}^k  \right|^2 ,\\
    f({\bf{tm}}^k)= {\bf F}^H{\bf F},
\end{align}
where ${\bf F}$ is a vector. According to the chain rule, the first derivative of the cost function can be written as,
\begin{align}
  {\frac{\partial f} {\partial {\bf tm }^k}} &= - {\frac{\partial f} {\partial {\bf F}}} {\frac{\partial {\bf F}} {\partial {\bf tm }^k}} \nonumber \\
  &= - {\frac{\partial f} {\partial {\bf F}}} {\frac{\partial  \left| {\bf{Q}} {\bf{tm}}^k  \right|^2} {\partial {\bf tm }^k}} \nonumber \\
  &= -4 {\bf F}^H \text{diag} ( conj( {\bf{Q}} {\bf{tm}}^k)) {\bf{Q}}.
\end{align}
Thus, we have the Hermitian of the first derivative as
\begin{align}
{\frac{\partial f} {\partial {\bf tm }^k}}^H &= -4 {\bf Q }^H \text{diag} ( {\bf{Q}} {\bf{tm}}^k)  {\bf F} \nonumber \\
&= -4 {\bf Q }^H \text{diag} ( {\bf Q } {\bf tm }^k) ({\bf I }^k - \left| {\bf{Q}} {\bf{tm}}^k  \right|^2 ).
\end{align}

\section{Derivation of the first derivative in the algorithm of phase correction}
\label{app:Phasecorrection}
\setcounter{table}{0} % 将表格计数设置为0
\setcounter{figure}{0} % 将图片计数设置为0
\setcounter{equation}{0} % 将公式计数设置为0

The optimization of solving the phase offset from the defocus intensity images is rewritten as
\begin{align}
   \min \limits_{\phi}  g(\phi)=\sum \limits_{n} \|  {\bf{I}}_n^{z_d}- \left|   {{\bf A}_n} e^{j\phi} \right|^2 \|^2_2,
    \label{eq:apoptphase}
\end{align}
where $ {{\bf A}_n}= {\bf K}_2^H {\bf P}\text{diag}({\bf h}){\bf K}_1 \text{diag}({\bf c}_{n})$.
Next, we define,
\begin{align}
   {\bf G}_n= {\bf{I}}_n^{z_d}- \left| {{\bf A}_n} e^{j\phi} \right|^2 ,\\
   g_n= {\bf G}_n^H{\bf G}_n.
\end{align}
By using the chain rule, we have,
\begin{align}
  {\frac{\partial g_n} {\partial e^{j\phi} }} &= - {\frac{\partial g_n} {\partial  {\bf G}_n}}    {\frac{\partial {\bf G}_n} {\partial e^{j\phi} }} \nonumber \\
  &= -{\frac{\partial g_n} {\partial  {\bf G}_n}}   {\frac{\partial \left| {{\bf A}_n} e^{j\phi} \right|^2} {\partial e^{j\phi} } } \nonumber \\
  &= -4{\bf G}_n^H \text{diag} ( conj(  {{\bf A}_n} e^{j\phi} )) {{\bf A}_n}.
  \label{eq:dgde}
\end{align}

We can have 
\begin{align}
\frac{ \partial e^{j\phi}}{ \partial \phi} = \text{diag}( je^{j\phi}).
\label{eq:dedphi}
\end{align}
By combining Eq.~\ref{eq:dgde} and Eq.~\ref{eq:dedphi}, we can get
\begin{align}
{\frac{\partial g_n} {\partial \phi }} = \text{real}(-4{\bf G}_n^H \text{diag} ( conj(  {{\bf A}_n} e^{j\phi} )) {{\bf A}_n}\text{diag}( je^{j\phi})).
\end{align}
It is easy to obtain,
\begin{align}
{\frac{\partial g_n}{\partial \phi }}^H=  \text{real}(-4 \text{diag}(-je^{-j\phi}) {{\bf A}_n}^H \text{diag}({{\bf A}_n}e^{j\phi})(  {\bf{I}}_n^{z_d} - \left| {{\bf A}_n} e^{j\phi} \right|^2)).
\end{align}
So we have,
\begin{align}
{\frac{\partial g}{\partial \phi }}^H&=  \sum \limits_{n} {\frac{\partial g_n}{\partial \phi }}^H \nonumber \\
&=  \sum \limits_{n}  \text{real}(-4 \text{diag}(-je^{-j\phi}) {{\bf A}_n}^H \text{diag}({{\bf A}_n}e^{j\phi})(  {\bf{I}}_n^{z_d} - \left| {{\bf A}_n} e^{j\phi} \right|^2)).
\end{align}

\end{appendices}

\section*{Funding}
This work was supported by the National Natural Science Foundation of China (61735017, 62020106002, 62005250, T2293751, and T2293752), the National Key Basic Research Program of China (2021YFC2401403), and Major Scientific Research Project of Zhejiang Lab (2019MC0AD02).

%\section*{Disclosures}
%The authors declare no conflicts of interest. 
%\section*{Supplemental document}
%See Supplement 1 for supporting content. 
\dag These authors share the first authors of this paper.

\bibliographystyle{ieeetr}
\bibliography{references}  %%% Uncomment this line and comment out the ``thebibliography'' section below to use the external .bib file (using bibtex) .





% \end{thebibliography}

\end{document}